\documentclass[prd,preprintnumbers,amsmath,amssymb,nofootinbib,12pt]  {revtex4}

\usepackage{epsfig}
\usepackage{bm}
\usepackage{dcolumn}
\usepackage[usenames]{color}
\usepackage{epstopdf}
\usepackage{amsmath}
\usepackage{amssymb}
\usepackage{url}
\usepackage{subfigure}
\usepackage{textcomp}
\usepackage{graphicx}
\usepackage{bm}
\usepackage{dcolumn}
\usepackage[usenames]{color}

\usepackage{epstopdf}

\usepackage{graphicx}
\newcommand{\beq}{\begin{equation}}
\newcommand{\eeq}{\end{equation}}
\newcommand{\bea}{\begin{eqnarray}}
\newcommand{\eea}{\end{eqnarray}}
\newcommand{\bi}{\begin{itemize}}
\newcommand{\ei}{\end{itemize}}

\newcommand{\sign}{\text{sign}}
\def\beq{\begin{equation}}
\def\eeq{\end{equation}}
\def\bea{\begin{eqnarray}}
\def\eea{\end{eqnarray}}

\makeatletter
\newcommand*{\rom}[1]{\expandafter\@slowromancap\romannumeral #1@}
\makeatother

\begin{document}

\title{Two-Field Quintessential Higgs model and the Swampland}

\author{Mehdi Es-haghi$^{1,2}$}
\email{eshaghi249(AT)gmail.com}

\author{Moslem Zarei$^{3,4}$}
\email{m.zarei(AT)iut.ac.ir}

\author{Ahmad Sheykhi$^{1}$}
\email{asheykhi(AT)shirazu.ac.ir}

\affiliation{$^1$ Physics Department and Biruni Observatory, Shiraz University, Shiraz 71454, Iran}
\affiliation{$^2$ Department of Physics, Islamic Azad University, Najafabad Branch, Najafabad, Iran}
\affiliation{$^3$ Department  of  Physics,  Isfahan  University  of  Technology,  Isfahan  84156-83111, Iran}
\affiliation{$^4$ CRANet-Isfahan,  Isfahan  University  of  Technology,  84156-83111,  Iran}


\date{\today}

 \begin{abstract}

We study a two-field model where a quintessence field with an exponential potential $e^{-\beta\phi/M_P}$ is coupled to the Higgs field. It is claimed that this model is consistent with
the proposed Swampland conjecture. We check this claim by calculating its inflationary observables. Although, these observables are in good agreement with the latest CMB data, but we find an upper bound $\beta \lesssim 8\times 10^{-3}$ that strongly disfavors the Swampland conjecture.

\end{abstract}

\maketitle

\section{Introduction}

Cosmologists consider two accelerating eras of cosmic expansion history. The first is an inflationary era where a scalar field, called inflaton, with negative pressure and a slow-roll evolution, is responsible for accelerating expansion in the early universe \cite{Guth}. Based on the Planck papers \cite{Aghanim:2018eyx, Akrami:2018odb}, single field potentials with plateau-like shapes are the most favored inflationary models by Planck temperature, polarization, and lensing data combined with the BICEP2/Keck Array BK15 data. In between the single field inflationary models, the Higgs model \cite{Salopek,Bezrukov:2007ep} has the best consistency with the Planck 2018 data. It belongs to a general class of inflationary models where a scalar field has a non-minimal coupling to gravity \cite{Futamase:1987ua}.

The second accelerating era is the late time acceleration, discovered in 1998 \cite{Riess:1998cb}. The most accepted premise for explaining this discovery is the dark energy (DE) theory. In this theory, dark energy is assumed as an unknown form of energy responsible for the presently observed acceleration of the galaxies. The dark energy would need to be very homogeneous and very low-density. It also has negative pressure. There are two general proposals for dark energy. The first is the cosmological constant, $\Lambda$, added to the Einstein equation. This constant represents a constant energy density filling space homogeneously. Indeed, this proposal leads to Lambda cold dark matter ($\Lambda$CDM) cosmological model, which has excellent agreement with the cosmological data \cite{Aghanim:2018eyx}. Following the exotic nature of dark energy, a second DE scenario has been proposed as a dynamical scalar field with infinitesimal energy density, called quintessence, which its slow-roll evolution at the late time can accelerate the universe \cite{Copeland:1997et}. This scalar field avoids the extreme fine-tuning of the cosmological constant \cite{Tsujikawa:2013fta}. Although, $\Lambda$CDM model with dark energy equation of state $\omega_{\Lambda}=-1$ satisfies the cosmological data very well without invoking any other explanation for dark energy, if future cosmological experiments show a deviation from $\omega_{\Lambda}=-1$, then quintessence may help to explain this deviation \cite{Akrami:2017cir}. 

In recent years, some string theorists have attempted to study dark energy in the context of string Swampland \cite{Vafa:2005ui,Ooguri:2006in}. They have studied constraints imposed by two proposed Swampland conjectures on cosmology. The first conjecture, called swampland distance conjecture, says that the scalar field excursions have an upper bound as 
\bea
\frac{\Delta \phi}{M_P}< c \sim O(1)~,\label{000}
\eea
where $M_{\textrm{P}}$ is the reduced Planck mass.
The second conjecture, called swampland de Sitter (dS) conjecture, is that dark energy is a rolling quintessence field $\phi$ which it's potential $V_Q(\phi)$ should satisfy the universal Swampland conjecture \cite{Obied:2018sgi,Agrawal:2018own,Ooguri:2018wrx,Cicoli:2018kdo}
\bea
M_P\frac{\lvert \nabla_{\phi}V_{\textrm{Q}} \rvert}{V_{\textrm{Q}}}> c \sim O(1)~,\label{00}
\eea
where $\lvert \nabla_{\phi}V_{\textrm{Q}} \rvert$ is the norm of the gradient of $V_{\textrm{Q}}$. These conjectures can be used also for inflaton fields. For a comprehensive review, covering the range of ideas from the Weak Gravity Conjecture to the de Sitter conjecture, one can see \cite{Palti:2019pca}. However, the results of some quintessence and inflation models are in tension with one or both of these conjectures \cite{Heisenberg:2018yae,Akrami:2018ylq,Colgain:2019joh}. 

Although the quintessence avoids the extreme fine-tuning of $\Lambda$, it has another tuning problem, namely its initial condition. Therefore, some people belive that explaining inflation and dark energy in unified scenarios with a single dynamical scalar field may overcome the above problems, at least the fine-tuning of $\Lambda$ \cite{Peebles:1998qn, Akrami:2017cir, Dimopoulos:2017zvq, AresteSalo:2021lmp, Dimopoulos:2022rdp}. In these scenarios, named single field quintessential inflation, the scalar field needs two plateau shoulders with an extremely large difference in their heights. The scalar field slowly rolls down from its inflationary plateau to its quintessential ones. After inflation, rolling the field towards its large negative values freezes at some $\phi_{\textrm{F}}$. In the late time, where the density of the radiation and the matter drops significantly, and the dark energy eventually dominates the energy density of the universe, the scalar field starts rolling down again and acts as the quintessence field.

As an example, a famous potential for describing dark energy is $V_{\textrm{Q}}(\phi)=V_{\textrm{0}}\:e^{-\beta \phi/M_{\textrm{P}}}$, where $\beta$ is a dimensionless positive constant and $V_{\textrm{0}}$ is potential energy constant. Interestingly, substituting this potential in the conjecture \eqref{00} gives $\lvert \nabla_{\phi}V_{\textrm{Q}} \rvert/V_{\textrm{Q}}=c$ (with $c=\beta$). However, this potential supports the inflation for $c\ll 1$ that strongly disfavors the conjecture \eqref{00} with $c >1$ \cite{Akrami:2017cir,Agrawal:2018own,Akrami:2018ylq}. Therefore, one may consider another adequate field that acts as the inflaton. On the other hand, if later studies confirm the Swampland conjecture \eqref{00}, the single-field slow-roll models of inflation in the landscape will be ruled out, and multi-filed ones with steep potentials will be replaced \cite{Achucarro:2018vey}.

The above discussions may motivate moving to the multi-field inflation and dark energy models. For this purpose, a new scenario, called two-field quintessential inflation, assumes that from the early times, there has been a two-dimensional potential in which one field is responsible for the dark energy. In contrast, inflation is driven by another field \cite{Akrami:2017cir}. During inflation, the inflaton field is central to the universe's evolution, while the quintessence field is sub-dominant. When the inflation ends, the inflaton field falls to the quintessence valley. In this model, the reheating occurs due to oscillations of the inflaton field near the potential minimum. Because of the small potential slope along the quintessence field axis, the quintessence field is sub-dominant until the density of reheating products (radiation and matter) decreases significantly. Eventually, the quintessence field undergoes the slow-roll evolution, which leads to the late time accelerated expansion. 

One of the proposed multi-field models is a two-field scenario in which a quintessence $\phi$ has a special coupling to the Higgs boson \cite{Denef:2018etk}. It is claimed that this coupling can prevent contradictory of Higgs potential with the swampland conjecture \eqref{00} around its local maximum. Moreover, this coupling may remove Higgs instability during inflation \cite{Han:2018yrk}. The aim of this paper is checking the consistency of this two-field potential with the Swampland Conjecture \eqref{00} during inflation. To achieve this aim, we construct a two-field quintessential Higgs inflationary potential which is an extension of the two-field potential in \cite{Denef:2018etk}. Then, we compare the cosmological observables of the potential with the latest CMB data. Finally, we study the swampland conjecture \eqref{00} for this potential.

 The paper is organized as follows: In Sec. \rom{2}, we make the model and study its potential behavior in the Einstein frame during inflation. Next, we deal with the slow-roll evolution of the model and its inflationary observables in Sec. \rom{3}. In Sec. \rom{4}, we calculate the number of e-folds to the end of inflation, which depends on some quantities like the energy density of the model, the reheating temperature, and the equation of the state of reheating. We review a proposal for solving the Higgs instability problem in Sec. \rom{5} that puts a constraint on the $\beta$ parameter in the quintessence model. Sec. \rom{6} is devoted to studying the CMB bounds on the inflationary observables of the model. Finally, this paper concludes with a summary in Sec. \rom{7}.

\section{The model}

Before introducing the model that we want to study in this section, we have a very short review on the Higgs and quintessence fields, separately.

The Higgs potential is as the following form  
\beq
V_{\mathcal{H}}=\lambda(\lvert \mathcal{H} \lvert ^2-v^2)^2,  \label{V1}
\eeq
where $\mathcal{H}=\begin{pmatrix} 0 \\
h/\sqrt{2}+v
\end{pmatrix}$ is the Higgs field, $\lambda$ is the self-coupling constant, and $v$ is the Higgs vacuum. The Higgs field is accompanied by a fundamental particle known as the Higgs boson, which was discovered in 2012 at the Large Hadron Collider (LHC) lab \cite{Chatrchyan:2012ufa}. Although the Higgs model plays an essential role in the standard particle physics (SM) model, it does not work well in describing inflation. That is because the large self-coupling $\lambda$ gives matter fluctuations larger than the observation. Fortunately, one can solve this problem by coupling the Higgs field non-minimally to gravity \cite{Bezrukov:2007ep}. This solution provides an inflation model that, on the one hand, has a root in the SM theory and, on the other hand, has excellent consistency with the CMB data \cite{Akrami:2018odb}. 

One of the well-known quintessence models for describing the late time dark energy is the model with the following potential
\beq
V_{\textrm{Q}}(\phi)=V_{\textrm{0}} e^{-\beta \phi/M_{\textrm{P}}}~,  \label{V1-0}
\eeq
where $\beta$ is a dimensionless positive constant and $V_{\textrm{0}}$ is potential energy constant. It is shown that for this model, the scale factor of the expanding universe grows as $a\sim t^{2/\beta^2}$. To provide an accelerating expansion, one should set $\beta<\sqrt{2}$ \cite{Liddle:1989mt,Kallosh:2003mt}.

The Higgs potential has a global minimum at $\left\vert\mathcal{H}\right\vert=v$ and a local maximum at $\left\vert\mathcal{H}\right\vert=0$. In the neighborhood of its local maximum, one obtains $\lvert \nabla_{\textrm{h}}V_{\mathcal{H}}(h) \rvert \sim 0$ and hence the conjecture \eqref{00} is violated \cite{Denef:2018etk}. Supposing the Higgs and the quintessence as the only scalar fields at the Electroweak (EW) scale \cite{Denef:2018etk, Murayama:2018lie}, a simple combination of \eqref{V1} and \eqref{V1-0} as 
\beq
V(h, \phi)=V_{\mathcal{H}}(h)+V_{\textrm{Q}}(\phi)~,  \label{vvv}
\eeq
gives a non-vanishing $\lvert \nabla V \rvert$. For this potential, one finds
\bea
M_P\frac{\lvert \nabla V \rvert}{V} \sim 10^{-55}~,\label{vvvv}
\eea
which is still in significant tension with \eqref{00}. A special combination of \eqref{V1} and \eqref{V1-0}, which may remove this tension, is 
\beq
V(h, \phi)=e^{-\beta(\phi-\phi_{\textrm{0}})/M_{\textrm{P}}}(V_{\mathcal{H}}(h)+\Lambda)~,  \label{V3}
\eeq
where $\phi_{\textrm{0}}$ is the value of $\phi$ at the present time \cite{Denef:2018etk}. In this model, one assumes a trilinear coupling $\dfrac{v^2}{M_{\textrm{P}}} \phi h^2$ between the Higgs and quintessence fields in the early universe. 

To study the consistency of the above model with swampland dS conjecture during inflation, first, we use \eqref{V3} to construct a two-field quintessential-Higgs inflation model. To this end, we use a generalized action in the Jordan frame for two scalar fields non-minimally coupled to gravity as follows \cite{Starobinsky:2001xq}
\bea
  S_J= \int d^4 x  \sqrt{-g} \left [ \frac{1}{2}M_{\textrm{P}}^2~[1+f(h,\phi)]R - \frac{1}{2} \partial_\mu h
\,  \partial^\mu h- \frac{1}{2} \partial_\mu \phi
\,  \partial^\mu \phi - V(h,\phi) \right]~,\label{J-action0-0}
\eea
where $R$ is the Ricci scalar and $f(h,\phi)$ is a non-minimal coupling term. The metric signature of this action is $(-, +, +, +)$.

To achieve a canonical form for the action \eqref{J-action0-0}, we use the following conformal transformation
\beq
\hat{g}_{\mu\nu}=\Omega^{2}\,g_{\mu\nu}~,  \label{O2}
\eeq
with the conformal factor, $\Omega^2$, defined as
\bea
\Omega^2=1+f(h,\phi)~, \label{O1}
\eea
where the non-minimal coupling $f(h,\phi)$ is a polynomial function.
Here, we assume that the quintessence field has minimal coupling to gravity. Therefore, the function $f(h,\phi)$ is only a function of the Higgs field, $f(h,\phi)=f(h)$. We choose a quadratic form for \eqref{O1} as 
\bea
\Omega^2 =1+\xi \frac{h^2}{M_{\textrm{P}}^2}~,  \label{O3}
\eea
where $\xi$ is the coupling constant between the Higgs field and gravity, to have a non-minimal model which is compatible with the particle physics and inflation scenario simultaneously, we need $1 \ll\sqrt{\xi}\lll 10^{17}$ \cite{Bezrukov:2007ep}.

The conformal factor \eqref{O3} leads to a non-minimal kinetic term for the Higgs field in \eqref{J-action0-0} \cite{Bezrukov:2007ep,Kallosh:2013tua}. To find a conformal form for this kinetic term, we assume that
\beq
\label{J-action2}
\dfrac{d\chi}{d h}=\sqrt{\dfrac{1}{\Omega^{2}}+6\left(\dfrac{\Omega'}{\Omega}\right)^2}~,
\eeq
where $\chi$ is the canonically normalized form of $h$ and the prime denotes derivative with respect to $h$ \cite{Bezrukov:2007ep}. Finally, using \eqref{O3} and \eqref{J-action2}, one obtains the following well-known form for the action \eqref{J-action0-0} in the Einstein frame 
\beq
\label{action2} S_{E}= \int
d^4 x  \sqrt{-\hat{g}} \left [ \frac{1}{2}M_{\textrm{P}}^2\hat{R} -\frac{1}{2} \partial_\mu \chi
\, \partial^\mu \chi -\frac{1}{2} e^{2 b} \partial_\mu \phi
\, \partial^\mu \phi - \hat{V}(\chi, \phi) \right]~,
\eeq
with
\bea \label{E2}
\hat{V}(\chi, \phi)=\frac{V(h,\phi)}{\Omega^4}=\frac{e^{-\beta(\phi-\phi_{\textrm{0}})/M_{\textrm{P}}} \left(\frac{\lambda}{4}( h^2-v^2)^2+\Lambda \right)}{\left(1+\frac{\xi}{M_{\textrm{P}}^2}h^2\right)^2}~,
\eea
and
\bea \label{E2-00}
b=-\frac{1}{2}\ln (\Omega^2)~.
\eea
 
For large field values of $h \gg M_{\textrm{P}}/\sqrt{\xi}$ or equivalently for $\chi \gg \sqrt{6}M_{\textrm{P}}$, the solution of \eqref{J-action2} can be approximated as
\bea \label{E3}
h=\frac{M_{\textrm{P}}}{\sqrt{\xi}} \exp \left(\frac{\chi}{\sqrt{6}M_{\textrm{P}}}\right)~.
\eea
Therefore, by substituting \eqref{E3} into \eqref{E2}, we find a product-separable potential in the limit $h\gg v \gg \Lambda$
\beq
\hat{V}(\phi, \chi)=\hat{V}(\phi)\hat{V}(\chi)~,\label{V4}
\eeq
where 
\bea \label{E5-1}
  \hat{V}(\phi)= e^{-\beta(\phi-\phi_{\textrm{0}})/M_{\textrm{P}}}~,
\eea
and
\bea \label{E5-1}
\hat{V}(\chi)= \frac{\lambda}{4}\frac{M_{\textrm{P}}^4}{\xi^2}\left(1+e^{-\sqrt{\frac{2}{3}}\chi /M_{\textrm{P}}}\right)^{-2}~.
\eea
Inflation is driven when the inflaton field with nearly flat potential rolls down very slowly compared to the expansion of the Universe \cite{Linde:2014nna}. At large field values $\chi \gg \sqrt{6}M_{\textrm{P}}$ and for small values of $\beta$, the potential \eqref{V4} is extremely flat. Therefore, inflation occurs via slow-roll evolution of the $\chi$ field. As the $\chi$ field is super-Planckian in this era, the model is contradictory with the swampland conjecture \eqref{000} which is the first problem of the model.

Approximating \eqref{O3} and \eqref{J-action2} for small field values $h \ll M_{\textrm{P}}/\sqrt{\xi}$ or equivalently for $\chi \ll \sqrt{6}M_{\textrm{P}}$, one finds $\Omega^2\simeq 1$ and $\chi \simeq h$. In this limit, the two-field potential \eqref{E2} approaches the potential \eqref{V3}, which is applicable for the eras after the end of inflation and we don't study it here.

\section{Slow-roll Inflation And Inflationary Observables}

The dynamics of the two scalar fields $\chi$ and $\phi$ during inflation is described by the Klein-Gordon and Friedmann equations as follows
\bea \label{E4}
&&\ddot{\chi}+3H\dot{\chi}+\hat{V}_{,\chi}(\phi, \chi)=b_{,\chi}e^{2b}\dot{\phi}^2~,\\\label{E4-0}
&&\ddot{\phi}+(3H+2b_{,\chi}\dot{\chi})\dot{\phi}+e^{-2b}\hat{V}_{,\phi}(\phi, \chi)=0~,\\\label{E4-1}
&&H^{2} = \frac{1}{3M_{\textrm{P}}^2}\left(\frac{1}{2}\dot{\chi}^{2}+\frac{1}{2}e^{2b}\dot{\phi}^{2}+\hat{V}(\phi, \chi)\right)~,\\\label{E4-11}
&&\dot{H}=-\frac{1}{2M_{\textrm{P}}^2}(\dot{\chi}^2+e^{2b}\dot{\phi}^{2})~.
\eea
where $H$ is the Hubble parameter, and dot and subscript comma denote partial derivative concerning time and scalar fields, respectively.

During the inflation era, $H$ is nearly constant, which corresponds to the condition
\bea \label{E4-2}
\epsilon^H \equiv -\frac{\dot{H}}{H^2}\ll 1~,
\eea
where $\epsilon^H$ is known as the Hubble slow-roll parameter. In this era, the potential energy of the model dominates over the kinetic energy as
\bea \label{E4-3}
\dot{\chi}^{2} \ll \hat{V}(\phi, \chi)~ 
  ,\:\:\:\:\: e^{2b}\dot{\phi}^{2} \ll \hat{V}(\phi, \chi)~.
\eea
Moreover, the scalar fields vary slowly during the inflation phase if
\bea \label{E4-4}
\lvert \ddot{\chi} \lvert ~\ll \lvert 3H\dot{\chi} \lvert ~ 
  ,\:\:\:\:\: \lvert b_{,\chi}e^{2b}\dot{\phi}^2 \lvert ~\ll \lvert 3H\dot{\chi} \lvert ~,
\eea
and 
\bea \label{E4-5}
\lvert \ddot{\phi} \lvert ~\ll \lvert 3H \dot{\phi} \lvert ~ 
  ,\:\:\:\:\: \lvert b_{,\chi}\dot{\chi}\dot{\phi} \lvert ~\ll \lvert 3H\dot{\phi} \lvert ~.
\eea

Equivalently, the slow-roll conditions for a single field $\varphi$ with a nearly flat inflationary potential $V(\varphi)$ are as
\bea \label{EEE}
\epsilon^{\varphi} \equiv \frac{M_P^2}{2}\left(\frac{V_{,\varphi}(\varphi)}{V(\varphi)}\right)^{2} \ll 1~ 
  ,\:\:\:\:\:\lvert \eta^{\varphi} \lvert  \: \equiv \lvert M_{\textrm{P}}^2\frac{V_{,\varphi \varphi}(\varphi)}{V(\varphi)}\lvert  \: \ll 1 ~.
\eea

For the two-field model, the slow-roll conditions become as the following
\bea \label{E5-11}
&&\epsilon^{\chi} \equiv \frac{M_{\textrm{P}}^2}{2}\left(\frac{\hat{V}_{,\chi}(\phi, \chi)}{\hat{V}(\phi, \chi)}\right)^{2}\ll 1~ 
  ,\:\:\:\:\:\:\:\:\:\:\:\:\:\:\eta^{\chi} \equiv M_{\textrm{P}}^2\frac{\hat{V}_{,\chi \chi}(\phi, \chi)}{\hat{V}(\phi, \chi)}\ll 1~,\\\label{E5-2}
&&\epsilon^{\phi} \equiv \frac{M_{\textrm{P}}^2}{2}\left(\frac{\hat{V}_{,\phi}(\phi, \chi)}{\hat{V}(\phi, \chi)}\right)^{2}e^{-2b}\ll 1~
  ,\:\:\:\:\:\eta^{\phi} \equiv M_{\textrm{P}}^2\frac{\hat{V}_{,\phi \phi}(\phi, \chi)}{\hat{V}(\phi, \chi)} e^{-2b}\ll 1~.
\eea
Substituting \eqref{V4} in \eqref{E5-11} and \eqref{E5-2}, the slow roll conditions take the forms as 
\bea \label{E5-111}
\epsilon^{\chi}= \frac{4}{3}\left(1-e^{\sqrt{\frac{2}{3}}\chi /M_{\textrm{P}}}\right)^{-2} \ll 1~
  ,\:\:\:\:\:\eta^{\chi}= \frac{4}{3}e^{-\sqrt{\frac{2}{3}}\chi /M_{\textrm{P}}}\frac{\left(-1+2e^{-\sqrt{\frac{2}{3}}\chi /M_{\textrm{P}}}\right)}{\left(1-e^{-\sqrt{\frac{2}{3}}\chi /M_{\textrm{P}}}\right)^{2}} \ll 1~,
\eea
and  
\bea \label{E5-22}
\epsilon^{\phi}= \frac{1}{2}\beta^2 \left(1+e^{\sqrt{\frac{2}{3}}\chi /M_{\textrm{P}}}\right)\ll 1~,
  \:\:\:\:\:\eta^{\phi}= \beta^2 \left(1+e^{\sqrt{\frac{2}{3}}\chi /M_{\textrm{P}}}\right)\ll 1~.\:\:\:\:\:\:\:\:\:\:\:\:\:\:\:\:\:\:\:\:\:\:\:\:\:\:\:\:\:
\eea

Inflation ends when the slow-roll conditions \eqref{E5-111} are broken as
\bea \label{End}
\epsilon^{\chi}=\frac{4}{3}\left(1-e^{\sqrt{\frac{2}{3}}\chi_{e} /M_{\textrm{P}}}\right)^{-2} \simeq 1~, 
\eea
where $\chi_{\textrm{e}}$ is the value of the inflaton field at the end of inflation. Using \eqref{End}, one can calculate $\chi_{\textrm{e}}$ numerically as 
\bea \label{E110}
\chi_{\textrm{e}}\simeq 0.94\:M_{\textrm{P}}~.
\eea

Imposing potential slow-roll conditions \eqref{E4-3}-\eqref{E4-5}, the equations \eqref{E4}-\eqref{E4-11} are simplified as the following forms
\bea \label{E6}
&&3H\dot{\chi}+\hat{V}_{,\chi}(\phi, \chi)\simeq 0~,\\\label{E6-0}
&&3H\dot{\phi}+e^{-2b}\hat{V}_{,\phi}(\phi, \chi)\simeq 0~,\\\label{E6-0-0}
&&H^{2} \simeq \frac{1}{3M_{\textrm{P}}^2}\hat{V}(\phi, \chi)~,\\\label{E6-0-00}
&&\dot{H}=-\frac{1}{2M_{\textrm{P}}^2}(\dot{\chi}^2+e^{2b}\dot{\phi}^{2})~.
\eea

Substituting \eqref{E6-0-0} and \eqref{E6-0-00} into \eqref{E4-2}, one obtains
\bea \label{E6-0-11}
\epsilon^H=\frac{3}{2}\frac{\dot{\chi}^2+e^{2b}\dot{\phi}^{2}}{\hat{V}}~.
\eea
By deriving $\dot{\chi}$ and $\dot{\phi}$ from \eqref{E6} and \eqref{E6-0} and substituting them into \eqref{E6-0-11}, we obtain the relation between the Hubble and potential slow-roll parameters of the model as
\bea \label{E6-1}
\epsilon^H=\epsilon^{\chi}+\epsilon^{\phi}~.
\eea

Substituting $\epsilon^{\chi}$ and $\epsilon^{\phi}$ from \eqref{E5-111} and \eqref{E5-22} into \eqref{E6-1}, we find
\bea \label{EEE}
\epsilon^H=\frac{4}{3}\left(1-e^{\sqrt{\frac{2}{3}}\chi /M_{\textrm{P}}}\right)^{-2}+\frac{1}{2}\beta^2 \left(1+e^{\sqrt{\frac{2}{3}}\chi /M_{\textrm{P}}}\right)~.
\eea
Later, we use this formula to calculate the inflationary observables of our model.

In the inflationary era, the quantum fluctuations of the scalar and gravitational fields produce scalar and tensor perturbations, respectively. The scalar spectral index defines the scale-dependence of the scalar power spectrum $\Delta_\zeta ^2$ 
\bea
n_\zeta -1=\frac{d\ln \Delta_\zeta ^2}{d\ln k}~,
\eea
and the normalized tensor-to-scalar ratio of the power spectrum is
\bea
r=\frac{\Delta_{\textrm{T}} ^2}{\Delta_\zeta ^2}~,
\eea
where $\Delta_{\textrm{T}} ^2$ determines the amplitude of gravitational waves. In the case of a two-field model with product-separable potential, the inflationary observables $\Delta_\zeta ^2$, $n_\zeta$ and $r$ have been calculated exactly in \cite{DiMarco:2005nq,Choi:2007su}. Therefore, we rewrite these observables for the potential \eqref{V4} as 
\bea \label{44}
&&\Delta_\zeta ^2 =\frac{\hat{V}}{24 \pi^2 M_{\textrm{P}}^4} e^{4(b_{\textrm{e}}-b_{\textrm{*}})}(\frac{u^2 \alpha^2}{\epsilon_{\textrm{*}}^{\chi}}+\frac{v^2}{\epsilon_{\textrm{*}}^{\phi}})~,\\ \label{E66}
&& n_\zeta -1=-2\epsilon_{\textrm{H}}-4\frac{e^{-4(b_{\textrm{e}}-b_{\textrm{*}})}}{u^2 \alpha^2/\epsilon_k^{\chi}+v^2/\epsilon_{\textrm{*}}^{\phi}}-\frac{1}{12}\frac{(\sqrt{\epsilon_{\textrm{*}}^{\phi}/\epsilon_{\textrm{*}}^{\chi}}u\alpha-\sqrt{\epsilon_{\textrm{*}}^{\chi}/\epsilon_{\textrm{*}}^{\phi}}v)^2}{u^2 \alpha^2/\epsilon_{\textrm{*}}^{\chi}+v^2/\epsilon_{\textrm{*}}^{\phi}}(\eta_{\textrm{*}}^b+2\epsilon_{\textrm{*}}^b) \nonumber\\
 && \:\:\:\:\:\:\:\:\ +2\frac{u^2\alpha^2 \eta_{\textrm{*}}^{\chi}/\epsilon_{\textrm{*}}^{\chi}+v^2\eta_{\textrm{*}}^{\phi}/\epsilon_{\textrm{*}}^{\phi}+4uv\alpha}{u^2 \alpha^2/ \epsilon_{\textrm{*}}^{\chi}+v^2 / \epsilon_{\textrm{*}}^{\phi}}+\frac{\sign (b_{,\chi})\sign (\hat{V}_{,\chi})v \sqrt{\epsilon_{\textrm{*}}^{\chi}\epsilon_{\textrm{*}}^b}}{u^2 \alpha^2/ \epsilon_{\textrm{*}}^{\chi}+v^2 / \epsilon_{\textrm{*}}^{\phi}}\left(\frac{v}{\epsilon_{\textrm{*}}^{\phi}}-2\frac{u}{\epsilon_{\textrm{*}}^{\chi}}\alpha\right)~,\\\label{E67}
&& r=\frac{8H_{\textrm{*}}^2}{(2\pi)^2 M_{\textrm{P}}^2 \Delta_\zeta ^2}=\frac{2\hat{V}}{3 \pi^2 M_{\textrm{P}}^4 \Delta_\zeta ^2}= 16 \frac{e^{-4(b_{\textrm{e}}-b_{\textrm{*}})}}{u^2 \alpha^2/\epsilon_{\textrm{*}}^{\chi}+v^2/\epsilon_{\textrm{*}}^{\phi}}~,
\eea
where 
\bea \label{E68}
&&\:\:\:\:\:\:\:\:\ u\equiv \frac{\epsilon^{\chi}_{\textrm{e}}}{\epsilon^{H}_{\textrm{e}}},\:\:\:\:\:\:\:\:\:\:\:\:\:\:\:\ v\equiv \frac{\epsilon^{\phi}_{\textrm{e}}}{\epsilon^{H}_{\textrm{e}}}~,\\
&&\:\:\:\:\:\:\:\:\ \epsilon^b\equiv 8 (b_{,\chi})^2,\:\:\:\:\:\:\:\ \eta^b\equiv 16 b_{,\chi \chi}~,\\
&&\alpha\equiv e^{-2(b_{\textrm{e}}-b_{\textrm{*}})}\left[1+\frac{\epsilon^{\phi}_{\textrm{e}}}{\epsilon^{\chi}_{\textrm{e}}}(1-e^{2(b_{\textrm{e}}-b_{\textrm{*}})})\right]~.
\eea

Planck data determine the observed scalar spectral index to be
\bea \label{FF}
n_\zeta=0.9649 \pm 0.0042~,
\eea
at 68\% CL \cite{Akrami:2018odb}. Moreover, the BICEP2, Keck Array and BICEP3 CMB polarization experiments put an upper limit on the observed tensor-to-scalar ratio as
\bea \label{HH}
r_{0.05} < 0.036~,
\eea
at 95\% CL \cite{BICEP:2021xfz}.

To calculate $n_\zeta$ and $r$, we need the values of $\chi$ and $\phi$ at the time of horizon crossing. In the next sections, we will calculate them. Then, we will compare $n_\zeta(\phi, \chi)$ and $r(\phi, \chi)$ with $\eqref{FF}$ and $\eqref{HH}$ to constraint the free parameters of the present model.

\section{Number of e-folds to the end of inflation}

The amount of inflation expansion during the time of horizon crossing, $t_*$ and the end of inflation, $t_{\textrm{e}}$, that is called the number of e-folds, $N_*$, is given in terms of the Hubble parameter H as $N_*=\int^{t_\textrm{e}}_{t_*}Hdt$, where the subscripts ``$*$'' and ``e'' indicate the value of quantities at the horizon exit and at the end of inflation, respectively. Assuming slow-roll conditions, one finds $N_*$ as a function of $\chi_*$ (or $\phi_*$) as
\beq
N_* =\frac{1}{M_{\textrm{P}}^2}\int^{\chi_*}_{\chi_\textrm{e}}\frac{\hat{V}(\chi)}{\hat{V}(\chi)_{,\chi}} d\chi ~. \label{E9}
\eeq
Substituting \eqref{E5-1} into \eqref{E9}, one finds
\beq
 N(\chi_*)= \frac{\sqrt{6}}{4 M_{\textrm{P}}}(\chi_{*}-\chi_{\textrm{e}})+\frac{3}{4}\left(e^{\sqrt{\frac{2}{3}}\frac{\chi_{*}}{M_{\textrm{P}}}}-e^{\sqrt{\frac{2}{3}}\frac{\chi_{\textrm{e}}}{M_{\textrm{P}}}}\right)~. \label{E10}
\eeq
By using the lower Lambert function, this equation can be inverted to \cite{Ellis:2015pla}
\beq
\chi_{*}(N_{*})\simeq \sqrt{\frac{3}{2}}M_{\textrm{P}} \ln \left[ \frac{4}{3}N_{*} - \sqrt{\frac{2}{3}} \frac{\chi_\textrm{e}}{M_{\textrm{P}}} + e^{\sqrt{\frac{2}{3}}\frac{\chi_{\textrm{e}}}{M_{\textrm{P}}}} \right]~. \label{E10-0}
\eeq
To determine $\chi_{*}$, one should calculate $N_{*}$. For this purpose, one may consider the connection between the time of horizon crossing of the observable cosmological scales and the time of their re-entering to the Hubble horizon as \cite{Liddle:2003as}
\bea
\frac{k}{a_{\textrm{0}}H_{\textrm{0}}}=\frac{a_{\textrm{*}}H_{\textrm{*}}}{a_{\textrm{*}}H_{\textrm{*}}}=e^{-N_{\textrm{*}}}\frac{a_{\textrm{e}}}{a_{\textrm{re}}}\frac{a_{\textrm{re}}}{a_{\textrm{eq}}} \frac{H_{\textrm{*}}}{H_{\textrm{eq}}}\frac{a_{\textrm{eq}}H_{\textrm{eq}}}{a_{\textrm{0}}H_{\textrm{0}}} ~ ,\label{E7}
\eea
where the comoving wavenumber $k$ equals the Hubble scale $a_* H_*$, and the subscripts refer to different eras, including the reheating (re), radiation-matter equality (eq), and the present time (0). By assuming the entropy conservation from the end of inflation to today and using the slow-roll condition in which $H^{2}_{*}\simeq V_*/3 M_{\textrm{P}}^2$, one obtains \citep{Martin:2010,Akrami:2018odb}
\bea
 N_*=67- \ln \left(\frac{k}{a_\textrm{0} H_\textrm{0}}\right)+\frac{1}{4}\ln \left(\frac{V_*^2}{M_{\textrm{P}}^4~\rho_{\textrm{e}}}\right)
 +\frac{1-3~\omega_{\textrm{int}}}{12(1+\omega_{\textrm{int}})}\ln\left(\frac{\rho_{\textrm{re}}}{\rho_{\textrm{e}}}\right)-\frac{1}{12}\ln g_{\textrm{re}} ~,\label{E8}
\eea
where $V_*$ is the potential energy of an inflationary model when $k$ leaves the Hubble horizon during inflation, $\rho_{\textrm{e}}$ and $\rho_{\textrm{re}}$ are the energy densities at the end of inflation and reheating, respectively, $\omega_{\textrm{int}}$ is the e-fold average of the equation of state between the end of inflation and the end of reheating and $g_{\textrm{re}}$ is the number of effective bosonic degrees of freedom at the end of reheating.

Now, we begin evaluating the quantities on the right-hand side of \eqref{E8} for the case of the present two-field model. In the third term of \eqref{E8}, $V_*=\hat{V}_*(\phi, \chi)$ can be calculated through the normalization of the power spectrum \cite{Baumann:2009ds} 
\bea \label{E11-0}
\Delta_{\zeta}^2 = \frac{k^3}{2\pi^2}P_{\zeta}(k)=\frac{1}{24\pi^2 M_{\textrm{P}}^4}\frac{\hat{V}_*(\phi, \chi)}{\epsilon^H} ~.
\eea
Therefore, one obtains 
\bea \label{E11-00}
\hat{V}_*(\phi, \chi)=24\pi^2 M_{\textrm{P}}^4~\Delta_{\zeta}^2~\epsilon^H~.
\eea
On the other hand, assuming zero acceleration at the end of inflation, $\ddot{a}_e=0$, one finds the condition $\dot{\chi}^2=\hat{V}_e(\phi, \chi)$ by solving the Friedmann equations. Using this condition, the energy density of the $\chi$ field at the end of inflation is given by
\bea \label{E13}
\rho_{\textrm{e}}=\frac{1}{2}\dot{\chi}^2+\hat{V}_{\textrm{e}}(\phi, \chi)=\frac{3}{2} \hat{V}_{\textrm{e}}(\phi, \chi)~.
\eea
 
In the fourth term of \eqref{E8}, $\rho_{\textrm{re}}$ is related to the reheating temperature $T_{re}$ through
\bea \label{E12}
\rho_{\textrm{re}}=\frac{\pi^2}{30} g_{\textrm{re}}T_{\textrm{re}}^4~.
\eea
Moreover, it is well known that $\omega_{int}$ of the reheating phase for large field models can be calculated from \cite{Turner}
\bea
\omega_{int}=\dfrac{p-2}{p+2}~, \label{E14}
\eea
where $p$ is the power of the inflaton field in the corresponding potential during the oscillating phase around its minimum. Using Taylor series expansion, one can approximate the inflationary potentials near their minimum as power law functions to study the behavior of these potentials during reheating. The Taylor expansion for the potential \eqref{V4} near its minimum gives
\bea
\hat{V}(\chi, \phi) \propto \chi^2~,
\eea
where
\bea \label{E2-2}
\chi \simeq \sqrt{\frac{3}{2}}\:\frac{\xi}{M_{\textrm{P}}}h^2~.
\eea
As a result, the behaviour of $\hat{V}(\chi, \phi)$ around its minimum as a large field potential $h^4$ (p=4) leads to
\bea \label{E2-222}
\omega_{\textrm{int}}=1/3~.
\eea
Therefore, the fourth term on right hand side of \eqref{E8} vanishes for $\omega_{int}=1/3$. As a consequence, deriving $\chi_{*}$ and $N_{*}$ for the present model is independent of reheating temperature.

Finally, by replacing \eqref{E10}, \eqref{E11-00}, \eqref{E13} and \eqref{E2-222} into \eqref{E8} we get 

\bea \label{E000}
&& \ln \left(\frac{k}{a_{\textrm{0}} H_{\textrm{0}}}\right)=67-\frac{\sqrt{6}}{4 M_{\textrm{P}}}(\chi_{*}-\chi_{\textrm{e}})+\frac{3}{4}\left(e^{\sqrt{\frac{2}{3}}\frac{\chi_{*}}{M_{\textrm{P}}}}-e^{\sqrt{\frac{2}{3}}\frac{\chi_{\textrm{e}}}{M_{\textrm{P}}}}\right)+\frac{1}{4}\ln \left(\frac{16\pi^2~\Delta_{\zeta}^2~\epsilon^H}{\hat{V}_{\textrm{e}}(\phi, \chi)}\right)
\nonumber\\
 &&\:\:\:\:\:\:\:\:\:\:\:\:\:\:\:\:\:\:\:\:\:\:\:\:\:\:\ -\frac{1}{12}\ln g_{\textrm{re}}~.
\eea
This equation is the basis for our subsequent analysis of the observables $n_\zeta$ and $r$, which are dependent on $\chi_{*}$. To obtain $\chi_{*}$ from \eqref{E000}, we need to know the values of the $\phi$ field and $\beta$ parameter during the inflation. In the next section, we study the bounds of these two values.

\section{Does Quintessence Save Higgs Instability?}

The Higgs inflation model suffers from Higgs instability since the quantum fluctuations of the Higgs field may have exceeded the instability EW scale during inflation \cite{Bezrukov:2007ep}. One proposal to overcome this problem is to couple a quintessence field $\phi$ to the Higgs field as the potential \eqref{V3} and define a new Higgs self coupling as $\tilde{\lambda}=\lambda e^{-\beta(\phi-\phi_{\textrm{0}})/M_{\textrm{P}}}$. Calculating the running of $\tilde{\lambda}$, one finds that the EW vacuum will be stable when \cite{Han:2018yrk}
\bea \label{G1}
e^{-\beta(\phi-\phi_{\textrm{0}})/M_{\textrm{P}}}\gtrsim 1.08~.
\eea
To find the upper limit for $e^{-\beta(\phi-\phi_0)/M_P}$ and the allowed range of $\beta$ that leads to Higgs stability, the authors of \cite{Han:2018yrk} studied the evolution of $\phi$ field from the era of inflation until today. Supposing ultra-slow-roll evolution of the $\phi$ field in the early universe and using the Klein-Gordon and Friedmann equations, they derived the initial condition of the $\phi$ field during inflation as a function of $\beta$. Comparing this initial condition with \eqref{G1}, they found a lower bound 
\bea \label{G2}
\beta>0.35\pm0.05~,
\eea
that leads to Higgs stability.

We will use \eqref{G1} in our subsequent calculations as one of the initial conditions to obtain $\chi_{*}$. However, we will check in the following whether the bound \eqref{G2} is consistent with the CMB constraint to $\beta$ or not.

\section{CMB Constraints to Quintessential Higgs Inflation observables}

Using the previous results and setting $g_{\textrm{re}}=1000$, $\Delta_{\zeta}^2=2\times 10^{-9}$ \cite{Aghanim:2018eyx}, $\chi_{\textrm{e}}\simeq 0.94\:M_P$ and $e^{-\beta(\phi-\phi_{\textrm{0}})/M_{\textrm{P}}} = 1$ \cite{Han:2018yrk}, \eqref{E000} is simplified as 
\bea \label{F00}
&& \ln \left(\frac{k}{a_{\textrm{0}} H_{\textrm{0}}}\right)=65.2-\frac{\sqrt{6}}{4M_{\textrm{P}}}\chi_{*}-\frac{3}{4}e^{\sqrt{\frac{2}{3}}\frac{\chi_{*}}{M_{\textrm{P}}}}+\frac{1}{4}\ln \left(\frac{4}{3}\left(1-e^{\sqrt{\frac{2}{3}}\frac{\chi_*}{M_{\textrm{P}}}}\right)^{-2}+\frac{\beta^2}{2}\left(1+e^{\sqrt{\frac{2}{3}}\frac{\chi_*}{M_{\textrm{P}}}}\right)\right) \nonumber\\
 && \:\:\:\:\:\:\:\:\:\:\:\:\:\:\:\:\:\:\:\:\:\:\:\:\:\:\:\: -\frac{1}{2}\ln \left(1-e^{-\sqrt{\frac{2}{3}}\frac{\chi_*}{M_{\textrm{P}}}}\right)~.
\eea
This is a relation between the wave number $k$ and $\chi_{*}$. Hence, by setting $\beta$ in \eqref{F00} as a free parameter and obtaining $\chi_*$ as a function of $\beta$ for pivot scale $k=0.002$ Mpc$^{-1}$, it is easy to calculate the observables \eqref{E66} and \eqref{E67} numerically.

\begin{figure}[h]
\epsfig{figure=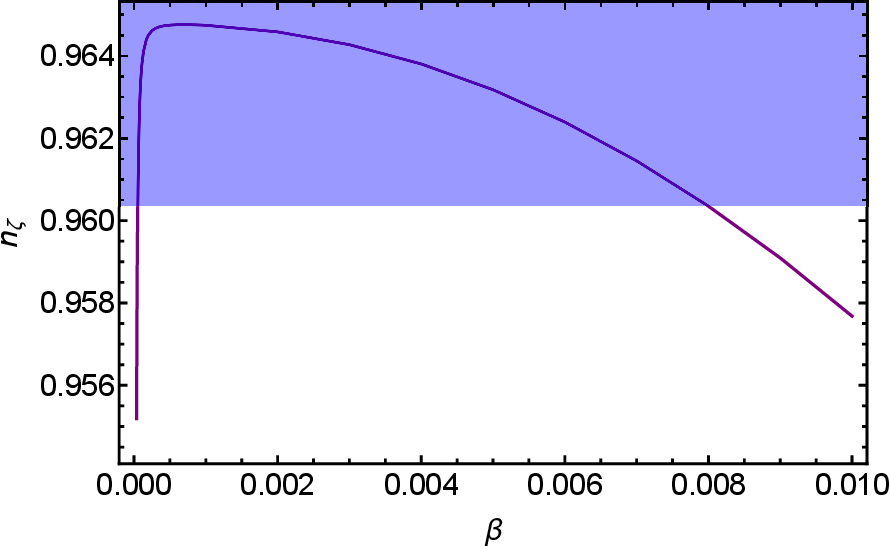,width=7.75cm}\hspace{2mm}
\epsfig{figure=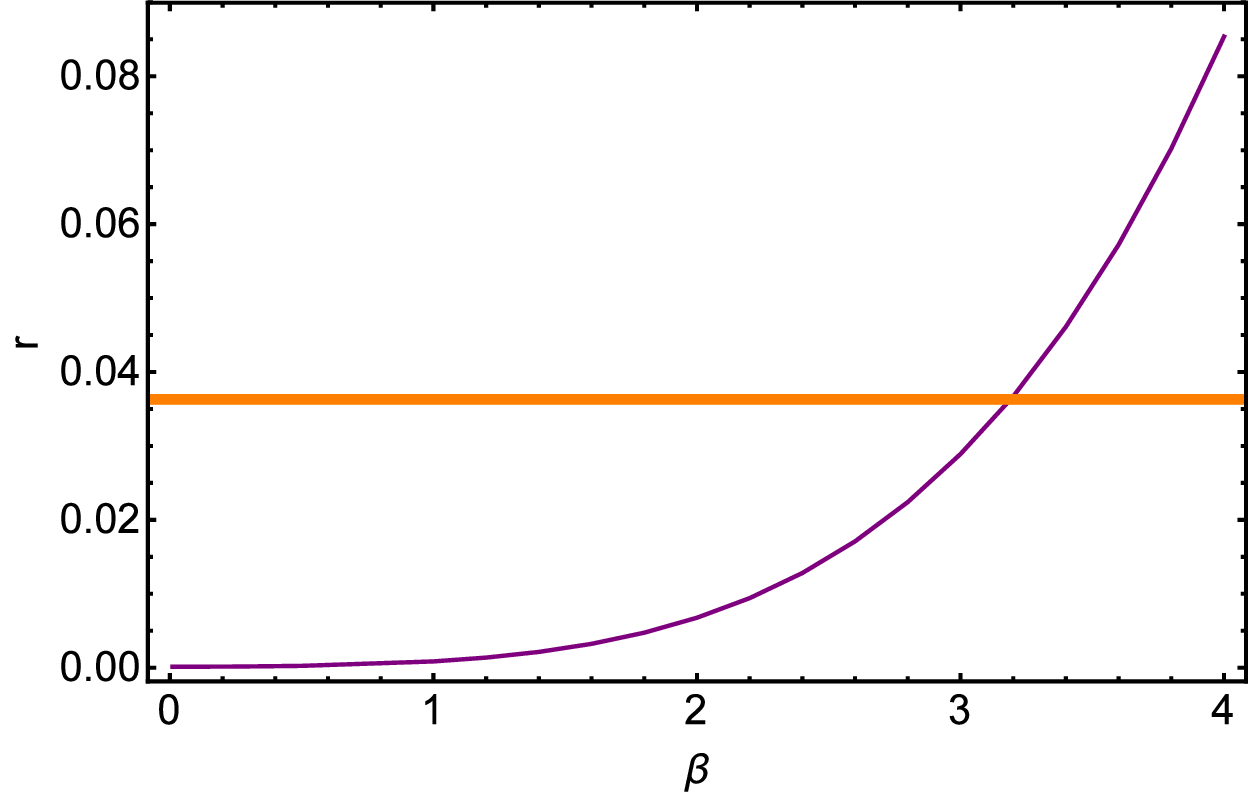,width=7.75cm}\hspace{2mm}
\caption{The spectral index, $n_\zeta$ (left panel) and the tensor-to-scalar ratio, $r$ (right panel) against $\beta$. The blue band in the left panel represents the 68\% CL region of the Planck data combined with the BICEP2/Keck Array BK15 data \cite{Akrami:2018odb}. The orange line in the right panel is due to the upper bound on $r_{\textrm{0.05}} < 0.036 $ at 68\% CL \cite{BICEP:2021xfz}.}
\label{fig}
\end{figure}

 Figure \ref{fig} shows the dependence of $n_\zeta$ and $r$ to $\beta$ for the present model. To keep $r_{0.05} < 0.036$, the right panel gives an upper bound $\beta\lesssim 3.2$. From the left panel, the upper and lower bound on $\beta$, which keeps $n_\zeta$ within 68\% CL region, is
\bea
5\times 10^{-5} \lesssim \beta \lesssim 8\times 10^{-3}~. \label{F1}
\eea
Replacing $\chi_{\textrm{e}}$ and $\chi_*$ of the present model into \eqref{E10} and \eqref{E67}, we find numerically the permitted values of $N_*$ and $r$ as
\bea
&&58.70 \lesssim N_{*} \lesssim 59.35~,\:\:\:\:\:\:\:\:   \textrm{for}~\:\:\:\:\:\:\:   5\times 10^{-5} \lesssim \beta \lesssim 8\times 10^{-3}~,\\
&&0.0014 \lesssim r \lesssim 0.0015~,\:\:\:\:\:\:\:    \textrm{for}~\:\:\:\:\:\:\:5\times 10^{-5} \lesssim \beta \lesssim 8\times 10^{-3}~,\label{F2}\
\eea
which are in a very good agreement with CMB data. As a result, the CMB constraint \eqref{F1} rules out the bound \eqref{G2}. It means that not only the quintessence does not remove Higgs instability, but also the Swampland Conjecture \eqref{00} is not satisfied which is the opposite of the claim of the authors in \cite{Han:2018yrk}.

\section{Conclusion }

In this work, we studied a previous proposal of two-field potential as \eqref{V3} in which the Higgs and quintessence fields have a trilinear coupling $\dfrac{v^2}{M_{\textrm{P}}} \phi h^2$. The main goal of the authors of this proposal was removing contradictory of Higgs potential with the swampland dS conjecture around its local maximum. In another related work, it was claimed that a lower bound on the free parameter of the noted model, $\beta$ as \eqref{G2} can guarantee the Higgs stability during inflation. To investigate the validity of these claims during inflation, we made an inflationary version of the noted two-field potential in which the Higgs field has a non-minimal coupling with gravity in Jordan frame. Moving to the Einstein frame through conformal transformation, we found the plateau-like two-filed potential \eqref{V4}. Calculating and plotting the inflationary observables of this potential, $n_\zeta$ and $r$ against $\beta$ and comparing them with the latest observed values of these observables, we found $\beta \lesssim 8\times 10^{-3}$ that strongly disfavors the Swampland conjecture \eqref{00}. Besides, this bound rules out the bound \eqref{G2}, which means that the quintessence does not save the Higgs instability problem, and one may search for other proposals to solve it.

\section*{\small Acknowledgement}

M. Es-haghi would like to thank Y. Akrami for his helpful comments and valuable discussions during the completion of this paper. Iran Science Elites Federation funded this project.

  \end{document}